\documentclass[aps,prx,twocolumn,superscriptaddress,10pt]{revtex4-2}

\usepackage{multirow,amsthm,amssymb,amsbsy,amsmath,epsfig,epstopdf,bm}

\usepackage{graphicx}
\usepackage{booktabs}
\usepackage{verbatim}
\usepackage{dcolumn}
\usepackage{color}
\usepackage[dvipsnames]{xcolor}
 \usepackage[normalem]{ulem}
\makeatletter
\renewcommand\@make@capt@title[2]{%
 \@ifx@empty\float@link{\@firstofone}{\expandafter\href\expandafter{\float@link}}%
  {\textbf{#1}}\@caption@fignum@sep#2\quad
}%
\makeatother
\newcommand{\prom}[1]{\langle #1 \rangle}

\newcommand{\gl}[1]{\textcolor{blue}{#1}}

\usepackage{hyperref}

\usepackage{lineno}
\newcommand{\RemoveSpaces}[1]{%
  \begingroup
  \spaceskip=1sp
  \xspaceskip=1sp
  #1%
  \endgroup}

\renewcommand{\onlinecite}[1]{\RemoveSpaces{$[\text{\citenum{#1}]}$}}

\newcommand{\sz}[0]{\sigma^z}
\newcommand{\sx}[0]{\sigma^x}

\newcommand{\ket}[1]{|#1\rangle}
\newcommand{\bra}[1]{\langle#1|}

\setcitestyle{super} %Line for the superscript citation of Nature
\begin{document}

%\title{\sout{Exploring the} Role of coherence in \ro{many-body} Quantum Reservoir Computing}
\title{Role of coherence in many-body Quantum Reservoir Computing}

\author{Ana Palacios}
\email{ana.palacios@qilimanjaro.tech}
\affiliation{Qilimanjaro Quantum Tech, Carrer de Veneçuela, 74, Sant Martí, 08019, Barcelona, Spain}
\affiliation{Departament de F\'{i}sica Qu\`{a}ntica i Astrof\'{i}sica, Facultat de F\'{i}sica,
Universitat de Barcelona, E-08028 Barcelona, Spain}
\affiliation{Institut de Ci\`{e}ncies del Cosmos, Universitat de Barcelona,
ICCUB, Mart\'{i} i Franqu\`{e}s 1, E-08028 Barcelona, Spain.}

\author{Rodrigo Mart\'inez-Pe\~na}
\affiliation{Instituto de F\'{i}sica Interdisciplinar y Sistemas Complejos (IFISC, UIB-CSIC), Campus Universitat de les Illes Balears E-07122, Palma de Mallorca, Spain}
\affiliation{Donostia International Physics Center, Paseo Manuel de Lardizabal 4, E-20018 San Sebastián, Spain}

\author{Miguel C. Soriano}
\affiliation{Instituto de F\'{i}sica Interdisciplinar y Sistemas Complejos (IFISC, UIB-CSIC), Campus Universitat de les Illes Balears E-07122, Palma de Mallorca, Spain}

\author{Gian Luca Giorgi}
\affiliation{Instituto de F\'{i}sica Interdisciplinar y Sistemas Complejos (IFISC, UIB-CSIC), Campus Universitat de les Illes Balears E-07122, Palma de Mallorca, Spain}

\author{Roberta Zambrini}
\affiliation{Instituto de F\'{i}sica Interdisciplinar y Sistemas Complejos (IFISC, UIB-CSIC), Campus Universitat de les Illes Balears E-07122, Palma de Mallorca, Spain}

\date{\today}

%%%%%%%%%%%%%%%
\begin{abstract}		
%%%%%%%%%%%%%%%

Quantum Reservoir Computing (QRC) offers potential advantages over classical reservoir computing, including inherent processing of quantum inputs and a vast Hilbert space for state exploration. Yet, the relation between the performance of reservoirs based on complex and many-body quantum systems and non-classical state features is not established. Through an extensive analysis of QRC based on a transverse-field Ising model we show how different quantum effects, such as quantum coherence and correlations, contribute to improving the performance in temporal tasks,  as measured by the Information Processing Capacity. Additionally, we critically assess the impact of finite measurement resources and noise on the reservoir's dynamics in different regimes, quantifying the limited ability to exploit quantum effects for increasing damping and noise strengths. Our results reveal a monotonic relationship between reservoir performance and coherence, along with the importance of quantum effects in the ergodic regime.

\end{abstract}

\maketitle

%% start line numbers here
% \linenumbers

\section{Introduction}

Over the last few decades, we have witnessed a remarkable surge in technological advancements driven by the introduction of Machine Learning (ML) across various domains, spanning both within and beyond the realm of physics
\cite{10.5555/1162264,MLreview,mehta2019high}. Applications in diverse areas such as pattern recognition, image and speech processing, and temporal series prediction have emerged as highly efficient and cost-effective. 
In this evolving scenario, Reservoir Computing (RC) is attracting increasing attention, heralded for its potential in swift real-time and efficient data processing \cite{2007RCreview,RC,libroMiguel,10261_267586}. The integration of an RC processor directly into a larger machine, essentially making it an intrinsic component with memory, streamlines implementation challenges and mitigates potential information loss during transport away from the main device. This, combined with the potential inherent benefits of quantum systems, motivates the growing interest and significance of Quantum Reservoir Computing (QRC) \cite{Fujii_Nakajima_2017,Mujal2021,Nokkala2021,Martinez2021,Tran_2021, QuantumNeuromorphicComputing,spagnolo2022experimental}.

QRC provides at least two potential advantages when compared to classical RC. First, QRC possesses an inherent capability for processing quantum inputs \cite{nokkala2023online,nokkala2024retrieving}, a feature that aligns with the quantum nature of the computational paradigm. Furthermore, QRC exhibits an exponential size of the phase space relative to the physical size of the system \cite{Martinez2020,PhysRevResearch.4.033007,Dudas2023, sakurai_2022_QRCscale_free}. Although achieving this exponential scaling may come at the cost of requiring more precision in the measurement phase \cite{mujal2023time}, it can lead to overcoming the constraints imposed by the feature space limitations inherent in classical systems. Such a prospect holds the promise of advancements in a variety of quantum technologies.

%where dissipation can be transformed into a constructive resource \cite{PhysRevResearch.5.023057,sannia2022dissipation,Domingo2023}.

To fully evaluate the practical realization of these advantages, it is fundamental to analyze various factors, including the system's robustness in the presence of unavoidable noise, which is paramount in the NISQ era \cite{Preskill2018quantumcomputingin,NISQ}, as well as its ability to effectively exploit the large number of degrees of freedom provided by the Hilbert space \cite{Fujii_Nakajima_2017,Mujal2021},  inherently related to the richness of the quantum dynamics. As shown in \onlinecite{Martinez2021, xia2022reservoir}, the ergodic regime is better suited for QRC for tasks requiring some nonlinear memory, in contrast to the limiting effect of quantum many-body localization (MBL) in this aspect. In particular, in the MBL phase, the Hamiltonian eigenstates display a low amount of entanglement, obeying the so-called area law, which implies a reduced expressiveness of the dynamics, compared to the entanglement volume law observed in the ergodic phase \cite{abanin2019colloquium}.

In this paper, we aim to investigate the impact of quantumness on the system dynamics by establishing a correlation between reservoir performance, quantified by the Information Processing Capacity (IPC) measure, and the amount of quantum coherence and correlations generated. The role of coherence in non-equilibrium processes has been extensively studied in previous literature \gl{\cite{colloquim_coherence}}, for instance in quantum thermodynamics \cite{santos_role_2019, francica_role_2019, francica_quantum_2020,van_vu_finite-time_2022},  quantum simulations \cite{PRXQuantum.2.017003}, quantum metrology \cite{PhysRevLett.123.180504}, quantum biology \cite{Q_bio}, or 
quantum dynamical phases \cite{styliaris_quantum_2019,anand_quantum_2021}. 
However, the role of quantum coherence in quantum machine learning and specifically in  QRC is just beginning to be explored \onlinecite{xia_configured_2023} and enables us to unveil the potential in going beyond classical approaches.
In addition, we aim to critically examine the influence of inevitable noise on reservoir dynamics. 
On the one hand, any QRC map must be irreversible to achieve fading memory. This can make engineered noise as the enabling ingredient 
 to enhance computational capabilities \cite{PhysRevResearch.5.023057,sannia2022dissipation,rodrigo2}. 
 On the other hand, the role of natural noise on top of the RC map is yet to be fully understood \cite{Govia_2021, ChenNurdin,gotting2023exploring}. For example, in the pioneering work of Ref.~\onlinecite{Fujii_Nakajima_2017} the possible constructive role of dephasing was hinted in QRC based on an erase-and-write map.
 Here, we systematically analyze the impact of such noisy effects and
 construct a practical scenario in which finite resources, represented by measurement accuracy, are assumed to model realistic implementations \cite{mujal2023time,PhysRevApplied.20.014051,khan2021physical}.

The paper is structured as follows: in Section~\ref{sec:model} we present the  QRC scheme based on an erase-and-write map and its different dynamical regimes. In Section~\ref{sec:coherence_role_noise} we address undesired decoherence effects, leading to realistic deviations from unitary evolution, and study coherence and correlations as a function of noise strength. In Section~\ref{sec:observable_trajectories} we analyze the dynamics of observables for different regimes and types of decoherence, which provides important insights into the interpretation of the reservoir's performance, presented in Section~\ref{sec:IPC}. In Section~\ref{sec:corr_vs_performance} we explicitly benchmark the correlations of the system versus its performance for different decoherence intensities, and in Section~\ref{sec:discussion} we discuss our results and conclusions. Some details of the calculation of the relevant correlation and performance measures employed in this work are reported in Sections~\ref{sec:corr_measures} and \ref{sec:IPC_calculation_details}.

\section{\label{sec:results}Results}

\subsection{\label{sec:model} Model}

We address time series processing through a many-body reservoir consisting of a fully connected network with uniformly distributed random couplings $J_{ij} \in [0, 1]$ and local disordered fields $h_i = h + w_i$ with $w_i \in [-W, +W]$.
The Hamiltonian reads
\begin{equation} \label{eq:Ham_hW} 
    H = \sum_{i>j} J_{ij} \sigma^x_i \sigma^x_j + \sum_i h_i \sigma^z_i 
    \end{equation}   
where $i=1, ..., N$ and $\sigma_i^\alpha = I_1 \otimes ... \otimes I_{i-1} \otimes \sigma_i^\alpha \otimes I_{i+1} \otimes ... \otimes I_N$.
This model was introduced in the early literature about QRC \cite{Fujii_Nakajima_2017, Martinez2020, Martinez2021}  motivated by the rich and diverse dynamics it can induce.
The introduction of disorder indeed allows for the appearance of MBL, driving the system out of the ergodic phase, a transition experimentally shown in ion traps \cite{Monroe_W}. 
Depending on the regime, the state of the system exhibits different quantum features. This allows us to establish the relation between the states' quantumness, manifested e.g. in entanglement or quantum coherences, and the capacity to process information in QRC.

\begin{figure}
    \centering
    \includegraphics[width = 0.49\textwidth]{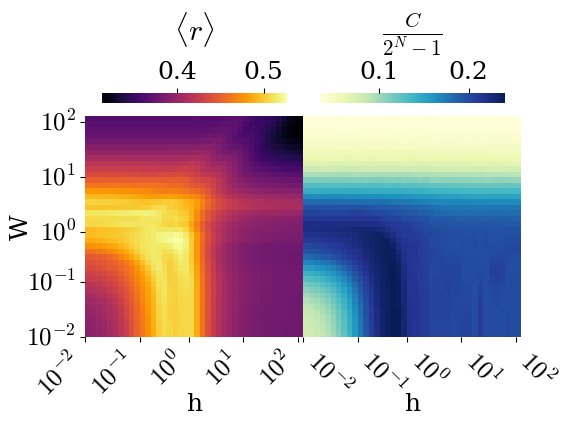}
    \caption{Average ratio between adjacent gaps $\prom{r}$ (left) and mean normalized coherence $C$ (in the z-basis) of the state of the QRC (right). Both quantities $\prom{r}$ and $C$ are averages over 100 configurations of disordered systems. 
    Coherence is evaluated for the QRC in the stationary regime after $\zeta=1000$ wash-out steps and
    with measures averaged over 100 consecutive time steps, where the input sequence was the same for all systems. 
    }
    \label{fig:model_heatmap_p0}
\end{figure}

In Fig.~\ref{fig:model_heatmap_p0} (left) the phase diagram of Eq.~\ref{eq:Ham_hW} is obtained addressing the average ratio between adjacent gaps $\prom{r}$, where $r_i = \frac{\min{(\omega_{n+1}, \omega_n)}}{\max{(\omega_{n+1}, \omega_n)}}$  and $\omega_n = E_n - E_{n-1}$, where $E_n$ are the ordered eigenvalues of $H$. The expected value of $\prom{r}$ in the ergodic phase, given by random matrix theory, is $\prom{r} \simeq 0.535$, while in the localized phase, $r$ follows a Poisson distribution and $\prom{r} \simeq 0.386$ \cite{atas_distribution_2013}.
We anticipate that we will consider temporal processing in QRC over significantly long series, analyze classical and quantum indicators,   and average over disordered configurations. Therefore, our analysis is performed on reservoirs of $N=5$ qubits. Changing the transverse field average and disorder values, in 
Fig.~\ref{fig:model_heatmap_p0} (left) we display an ergodic phase region (light color) that is progressively changing into a regime displaying localization (dark color). Due to the reduced size of the system, finite-size effects can limit a fully-fledged many-body localized phase for strong random qubit detuning. However, since the main features of MBL in the context of QRC are already observed in this regime, we will abuse language throughout this work and refer to it as the MBL region. As mentioned in the introduction, a link between the different dynamical phases of the system and the resulting reservoir's processing capacity has already been established in previous works \cite{Martinez2021, xia2022reservoir}. Intuitively, this connection can be understood from how well information propagates across the system, which allows for a better mixture and recombination of the input data through the reservoir map. Thus, higher nonlinearities and faster convergence of the processing capacities overall are expected in the ergodic regime. On the other hand, the opposite is expected in the MBL phase due to the large number of quasi-local integrals of motion that arise in this regime, which hinder information flow.

We now introduce the QRC operation, based on the out-of-equilibrium dynamics of the reservoir \eqref{eq:Ham_hW} driven by an erase-and-write map \cite{Fujii_Nakajima_2017}, a standard choice within the field. The reservoir map consists of injecting some random, uniformly distributed input $s_k \in [0, 1]$ and then letting the system evolve freely for a time $\Delta t$:
\begin{equation}
    \rho(k\Delta t) = 
    e^{-iH\Delta t} \rho^{\prime}((k-1)\Delta t) e^{iH\Delta t} 
    \label{eq:unitary_evol_Dt}
    %\label{eq:input_injection_map}
\end{equation}
where $\rho^{\prime}(k\Delta t) = \rho_1(s_k) \otimes \text{Tr}_1 \left[\rho(k\Delta t)\right]$, $H$ corresponds to Eq.~\eqref{eq:Ham_hW} and $\text{Tr}_1$ is the partial trace realized over the first qubit. The specific encoding of $s_k$ into $\rho_1$ will be discussed in the next section.
The erasure associated with each input injection leads to the dissipation that is actually needed for the proper QRC operation \cite{Fujii_Nakajima_2017,ChenNurdin, sannia2022dissipation,rodrigo2}.
 After a certain transient regime in which the above scheme is repeated for $\zeta$ expendable (possibly random) inputs, also referred to as wash-out, the reservoir reaches a stationary performance and no longer depends on its initial conditions, but its state becomes a function of the input history (echo state property). In the following, we will always focus on this operational regime.

To complete the QRC architecture, beyond the input and the reservoir, a third layer is introduced, i.e. the output, consisting in $M$ observable expectation values \cite{2007RCreview,Mujal2021} that form the data matrix $\hat{X}$. Since these expectation values stem directly from the simulation of the quantum system, we add some random Gaussian noise of mean 0 and standard deviation $\overline{\sigma} = 0.001$ to account for the effect of statistical noise that would be present in an experimental setting. This value of the standard deviation would correspond to a number of measurements of the order of $n_{\text{meas}}\sim 10^6$ as elaborated in \onlinecite{mujal2023time}. We choose this value because it is a realistic figure for the number of shots that may be employed for measurement in a high-precision experiment with superconducting qubits \cite{arute_quantum_2019, havlicek_supervised_2019}.
%assuming perfect readout fidelity, 
The bound on $n_{\text{meas}}$ is due to the fact that measurement is usually the slowest operation of the experiment ($\sim 0.1-1\mu$s for superconducting qubits), and thus dominates its total time.
With the addition of this Gaussian noise, the data matrix is thus transformed into $\Tilde{X}$, which is then used to train the output weights $W$ and the subsequent testing of the reservoir, as illustrated in Fig.~\ref{fig:model_schematic}. A comparison with what would be observed in the limit of performing an asymptotically large number of measurements is presented in Appendix~\ref{appendix:gaussian_smoothening}.

\subsection{\label{sec:coherence_role_noise}Coherences and the role of noise}

We now investigate whether there is a relationship between the localized and ergodic dynamical phases and the way in which the evolution of the reservoir induces quantum coherence.
In this work, we encode the $k$-th element of the time series in the populations of a classical state of the first qubit: $\rho_1(s_k)=\begin{pmatrix}
s_k & 0\\
0 & 1-s_k
\end{pmatrix}$. 
Therefore, quantum coherence $C$ (defined as the sum of the absolute values of the nondiagonal elements of the density matrix, which corresponds to the standard l1-norm of coherence, see Methods section~\ref{sec:methods}) can only be increased during the unitary step, and not at new input injections.
\begin{figure}
    \centering
    \includegraphics[width=0.39\textwidth]{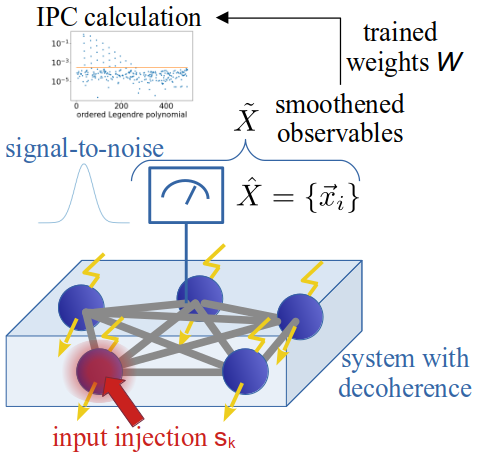}
    \caption{Illustration of the different layers of the simulation. The natural dynamics of the system with decoherence is modified by the input injection map. Measurement of the relevant set of observables is performed right before input injection, and ensemble measurements are considered. Once the observables are collected, we apply Gaussian noise over the collected expectation values to simulate signal-to-noise ratio limitations. The resulting data matrix $\tilde{X}$ is the one used for training and the calculation of the IPC contributions.}
    \label{fig:model_schematic}
\end{figure}

The average coherence $C$ with respect to the $z$ direction, for the stationary regime under the input injection map described in Eq. \ref{eq:Ham_hW}, is shown in Fig.~\ref{fig:model_heatmap_p0} (right), where we initialized the evolution in the maximally incoherent state. 
A side-by-side comparison between $\prom{r}$ and the coherence displays a clear relation between the ergodic phase and the build-up of the largest coherence. A similar relation has been reported in the Heisenberg model as well \cite{dhara2020quantum}. 
This points to the fact that correlations remain higher in this regime, regardless of the additional dynamics introduced by the input injection map, indeed due to the better flow of information facilitated by the underlying Hamiltonian of the system.
Different regimes will be explored in the following, along the line $h = 1$: $W=0$  is representative of the ergodic regime while for $W=10$ localization arises and the quantum coherence of the state is difficult to build up.

\begin{figure}
    \centering
    \includegraphics[width = 0.40\textwidth]{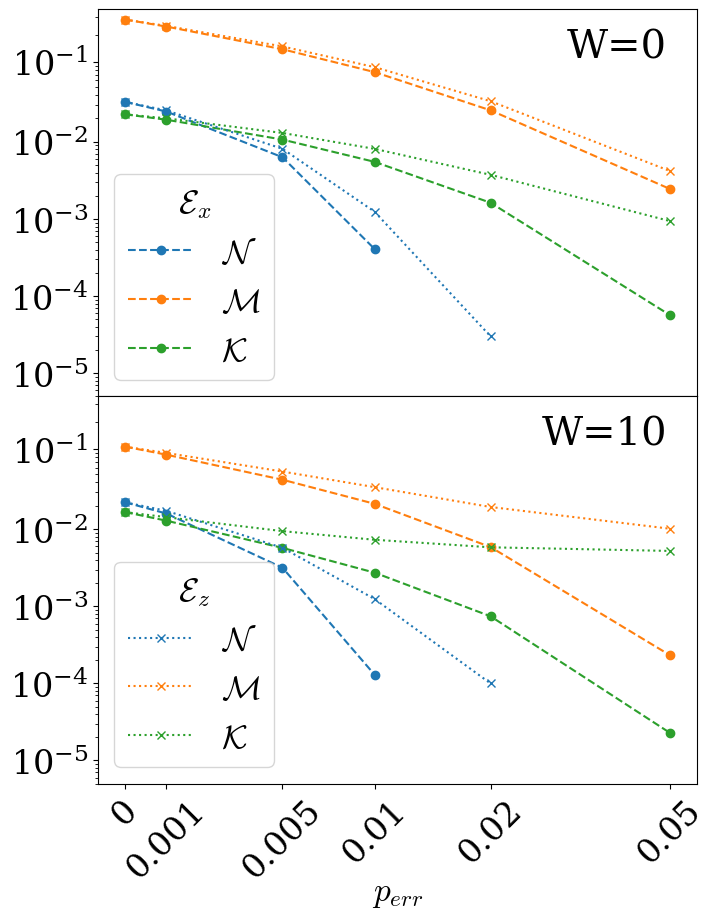}
    \caption{Analysis of correlations in the ergodic phase (upper plot) and in the MBL phase (lower plot). Coherence $C$ line omitted being undistinguishable from $\mathcal{M}$ (see main text). Points smaller than $10^{-6}$ have been omitted. Averages are taken over 100 measurements and 10 Hamiltonian realizations after $\zeta=500$ wash-out steps. The input sequence was the same for all systems and measurement runs at different noise intensities.}
    \label{fig:NMK_W0W10_noerrobars}
\end{figure}

While the presence of dissipation in QRC is necessary to guarantee fading memory, it is also clear that decoherence can eventually hinder the state's quantumness and the desired larger expressivity in the Hilbert space. On the other hand, several studies have shown that localization signatures can survive even in the presence of a bath \cite{nandkishore2014spectral,johri2015many,levi2016robustness,fischer2016dynamics}, which could lead to interesting dynamical effects in the QRC context. Overall, it has indeed been shown that decoherence can have competing effects \cite{Fujii_Nakajima_2017,sannia2022dissipation,gotting2023exploring}.  

The study of (tunable) decoherence in different directions ($x,y,z)$ provides an understanding of where the relevant correlations are stored and helps to understand the origin of the correlations that are responsible for the performance of the reservoir.
We address decoherence through a Markovian master equation, modeling either a bit flip or a phase flip channel \cite{NielsenChuang,ChenNurdin} acting on each qubit independently. This assumption leads to the following Lindblad equation for the evolution of the system:
\begin{equation}
    \dot{\rho} = -i[H,\rho] + \gamma \sum^N_{i=1}(\sigma_i^{\alpha} \rho \sigma_i^{\alpha} -\rho)
    \label{eq:lindblad}
\end{equation}
where $\alpha = x$ corresponds to the bit flip channel and $\alpha = z$ to phase flip. In our simulations, we implement this Lindbladian in a trotterized fashion as explained in Appendix~\ref{appendix:trotterization_lindblad}, which can be shown to correspond to the alternating application of the unitary dynamics for a shorter time $\delta t=\Delta t/\eta \to 0$ and the decoherence map with a given error strength $p_{err}$, which relates to the noise strength in Eq.~\eqref{eq:lindblad} as $p_{err} = \frac{1}{2}(1 - e^{-2\gamma \delta t})$. 
Looking at the QRC erase-and-write map in Eq.~\eqref{eq:unitary_evol_Dt}, this (uncontrolled) noise effect replaces the unitary (noiseless) evolution between input injections, as we represent in Fig.~\ref{fig:model_schematic}.

The effect of decoherence is analyzed on different correlation measures such as the quantum hookup $\mathcal{M}$ \cite{giorgi_hallmarking_2018}, which was defined as a measure of the total quantumness of a density matrix, including both nonlocal (quantum and classical) correlations and coherence. In the simulated instances of our system, $\mathcal{M}$ practically coincides with the mutual information (the difference being of the order of $10^{-6}$ or lower) and thus indicates the total correlations in the system. We will also consider the quantity $\mathcal{K}$ \cite{giorgi_hallmarking_2018}, the totally classical correlations, which correspond to the mutual information of the completely decohered state. Finally, we will also study the entanglement negativity $\mathcal{N}$ \cite{vidal2002} for a measure of purely quantum correlations. In order to keep the discussion on the results concise we refer the reader to the Methods section~\ref{sec:methods} for the mathematical definitions of each of these quantities. However, we highlight that the negativity and the mutual information employed here correspond to standard correlation measures in quantum information science. We further note that, according to Eq. (16) in \onlinecite{giorgi_hallmarking_2018}, $\mathcal{M} = C + \mathcal{K}$, which means that $\mathcal{M}$ and $C$ follow very similar trends for $\mathcal{K} \ll \mathcal{M}, {C}$, as it is generally the case throughout this study. Even though the measure of quantum correlations that can be related to $\mathcal{M}, C$ and $\mathcal{K}$ is quantum discord \cite{giorgi_hallmarking_2018}, we characterize them through the negativity instead in the interest of computational efficiency.

In Fig.~\ref{fig:NMK_W0W10_noerrobars} we plot these quantum and classical correlations for increasing bit flip and phase flip noise strengths. 
We observe that all the correlation indicators in the noiseless QRC are higher in the ergodic phase (for $W=0$) than in the localized one ($W=10$). 
On the other hand, in both ergodic and localized regimes, classical correlations are more robust to phase flip noise, as expected.
Different quantitative effects are appreciated for bit flip (dashed lines) and phase flip (dotted lines) noise as well; bit flip generally induces a faster decay of all indicators.
Interestingly, for $W=10$ phase flip noise barely affects the totally classical correlations, and its effect is generally milder than bit flip also in $\mathcal{M}$ and in $\mathcal{N}$. 
The greater impact of phase flip on $\mathcal{M}$ and on $C$ in the ergodic phase compared to the localized phase indicates that more information is being carried by the off-diagonal elements of the density matrix in the former. 

\subsection{\label{sec:observable_trajectories} Observable dynamics}

\begin{figure}
    \centering
    \includegraphics[width = 0.45\textwidth]{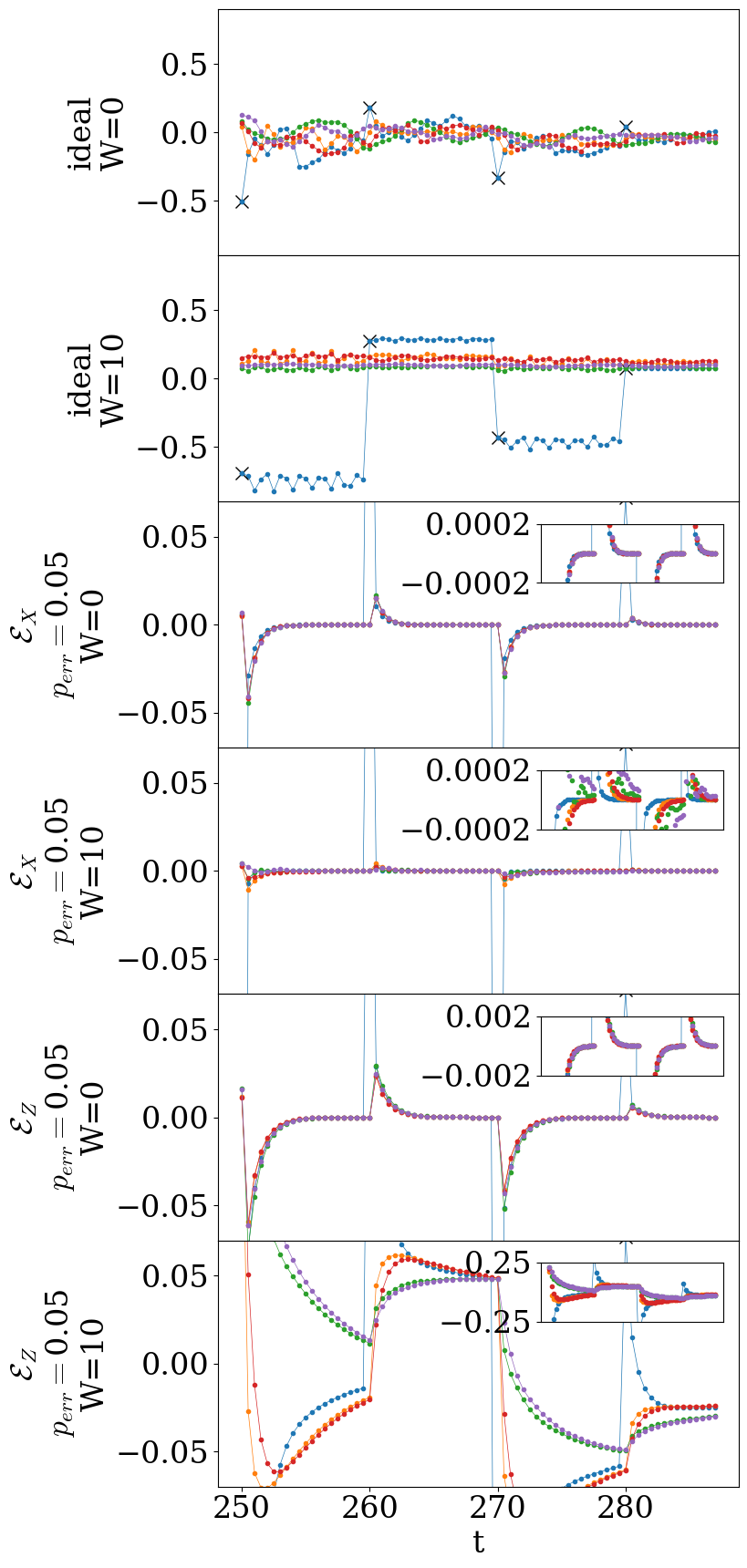}
    \caption{System response of all local Z-observables measured subjected to decoherence. Blue dots correspond to points on the trajectory of the first qubit's Z-observable $\prom{\sigma^z_1}$ (where input is injected), and the rest of the colors to the remaining Z-observables $\prom{\sigma^z_{i\neq 1}}$. Crosses correspond to points taken right after input injection.}
    \label{fig:Ztrajectories}
\end{figure}

%{\large[[PARAGRAPH ACCOMPANYING Fig.~\ref{fig:Ztrajectories}]]}
The performance in processing time series is ultimately related to the ability of the system to respond to each input injection in a distinctive way and display a dependence on previous inputs. 
In Fig.~\ref{fig:Ztrajectories} we show the dynamics of the expected values of the $\sz$ observables of each qubit during four input injections. Deeply different responses are found in the limiting cases of no noise (ideal) and strong bit flip and phase flip noise (for $p_{err}=0.05$). In the ideal case (unitary dynamics between input injections), a rather stiff qubit dynamics is shown in the MBL regime (W=10), not following the input qubit changes, as the input injection map only affects the first qubit and information does not flow to the rest of the system \cite{Martinez2021}.
On the other hand, in the ergodic case ($W=0$), a collective response to each new input is a signature of the better performance of this regime for information processing.

While this is consistent with previous observations \cite{Martinez2021} in the ideal erase-and-write map,  the presence of decoherence induces strong departures from this kind of dynamics.
The first observation is that in the presence of bit flip, the full evolution map (input injection plus noise) evolves towards a single fixed point, the maximally incoherent state (in the z direction) independently of the inputs. Thus for strong noise the dependence on the input is quickly lost. In the ergodic regime, the phase and bit flip maps have both a damaging effect, but a peculiar response is found in the presence of localization. Indeed, the stationary regime still retains some input dependence, even when the observable expectations are small and can become hardly useful in practical implementations, where statistical and experimental noise limits the precision (see  Section~\ref{sec:IPC}).

\subsection{\label{sec:IPC}Information Processing Capacity (IPC)}

A powerful indicator for the processing capacity in machine learning for time series processing is the Information Processing Capacity (IPC) measure, introduced in \onlinecite{Dambre}, first addressed in QRC in \onlinecite{Martinez2020}  and recently also related to the polynomial chaos expansion in \onlinecite{kubota2021unifying}. The IPC allows benchmarking performance in different regimes and its components (of different degrees) reflect the reservoir's ability to address $d$-degree nonlinearities.
This is done by evaluating the system's performance in a class of tasks that consist of approximating a certain polynomial target function, where the set of polynomials for each degree $d$ must form an orthonormal basis in function space. Thus, the target functions for the reservoir have the following formulation:
\begin{equation}
    \overline{y}_k = \prod_i \mathcal{P}_{d_i} [\Tilde{s}_{k-i}], \qquad \sum_i d_i = d,
    \label{eq:targetfuns_IPC}
\end{equation}
where the $\mathcal{P}_{d_i}$ appearing in Eq.~\eqref{eq:targetfuns_IPC} refers to the Legendre polynomial of degree $d_i$ (any orthogonal polynomials can be considered), and the $\Tilde{s}_{k-i}$ are the input with a delay of $i$ time steps, and where the relation between the task input $\Tilde{s}_k$ and the input injected in the quantum state $s_k$ is $\Tilde{s}_k = 2s_k - 1$. The fidelity with which the reservoir can perform one of these tasks is then a contribution to the degree capacitance $\mathcal{I}_{d}$ (see Section~\ref{sec:IPC_calculation_details} for further details on the calculation). Interestingly, the total capacity $\mathcal{I}_{tot}=\sum_{d=1}^{\infty} \mathcal{I}_d$ is bounded from above by the number of output functions $M$, and saturation of the bound implies that the system has fading memory:
\begin{equation}
    0 \le \mathcal{I}_{tot} = \sum_{d = 1}^{\infty} \mathcal{I}_d \le M
    \label{eq:IPC_bound}
\end{equation}
The output layer can be realized by accessing different sets of $M$ observables.  In our work, we will consider all possible $\langle \sz_i\sz_j\rangle $ pairs of the $i,j$ reservoir qubits (labeled as $ZZ$ set),  or combining $ZZ$ and local fields in $Z$ ($Z+ZZ$ set), or all possible pairs of observables laying outside of the computational axis ($XX + XY + YX + YY$ set). In the following, we will always consider the normalized IPC, i.e., $\mathcal{I}_{tot}/M \to \mathcal{I}_{tot}$, such that $0 \leq \mathcal{I}_{tot} \leq 1$.

Interestingly, as higher-order terms, some of these observables can have small expectation values, raising the question of their practical relevance in QRC implementations. To reach insightful conclusions, we model the effect of finite sampling and experimental imperfections with
Gaussian noise. As anticipated in Fig. \ref{fig:model_schematic}, the output layer is then obtained with tilde variables $\Tilde{X}$.
This leads to a more realistic signal-to-noise ratio (SNR) in view of experimental realizations. The detrimental effect of statistical noise was already pointed out in previous works \cite{Govia_2021,mujal2023time,polloreno2023limits,hu2023tackling,nokkala2021high} in connection to the scalability and expressivity of QRC. In particular, Refs.~\onlinecite{vettelschoss2021information,hu2023tackling,polloreno2023limits} show that the total capacity $\mathcal{I}_{tot}$ can decrease in the presence of statistical noise.
As it is further developed in Appendix~\ref{appendix:gaussian_smoothening}, not taking this into account can provide misleading results, especially in the MBL case, as it becomes apparent in the cases with $W=10$ and $\mathcal{E}_X$ in Fig.~\ref{fig:Ztrajectories}. 

\begin{figure}
    \centering
    \includegraphics[width=0.49\textwidth]{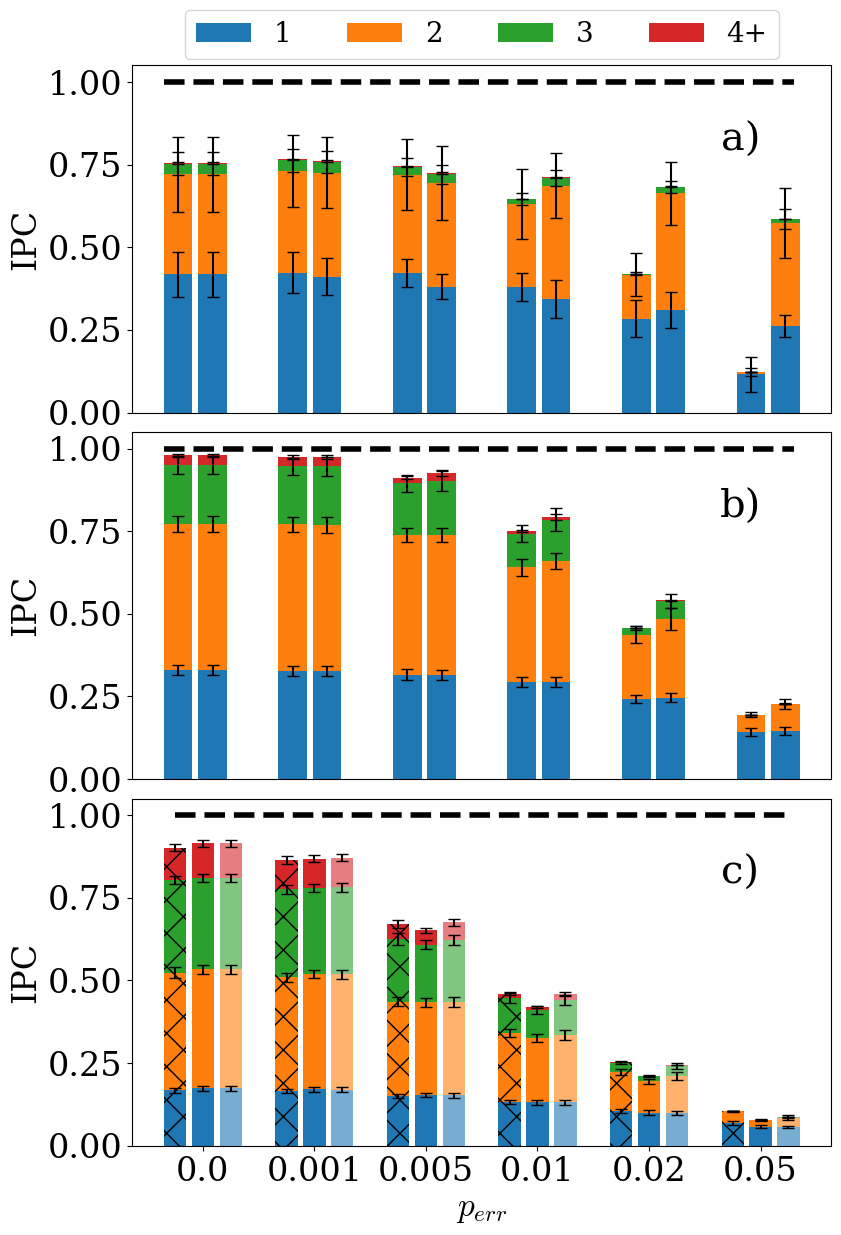}
    \caption{IPC contributions for a) $W=10$ and all $ZZ$ pairs as output functions; b) $W=0$ and the $Z+ZZ$ set of observables; and c) $W=0$ and multiplexed $ZZ$ set with $V=4$ (hatched, left) and $XX + XY + YX + YY$ set of observables subject to $\mathcal{E}_X$ (center) and to $\mathcal{E}_Z$ (right), which amounts to the same number of output functions. Left and right bars in a) and b) correspond to $\mathcal{E}_X$ and $\mathcal{E}_Z$ noise, respectively. 
    }
    \label{fig:IPC}
\end{figure}

In Fig.~\ref{fig:IPC} we show some representative examples of reservoir performance in different dynamical regimes and for different sets of observables. 
Motivated by the dynamics of the observables in Fig. \ref{fig:Ztrajectories}, we first investigate the influence of noise on the capacities for the localized phase.
Fig.~\ref{fig:IPC}a illustrates the performance of the MBL phase for the $ZZ$ observable set, where 
we find there is a greater robustness to phase flip than to bit flip noise. This aligns with the decoherence effects seen on correlations in Fig.~\ref{fig:NMK_W0W10_noerrobars}. In addition, we note that the IPC does not reach the maximum value even in the noiseless case, indicating that the localized regime is not computationally efficient. As a general trend, this regime exhibits relatively large error bars due to greater variability among individual random realizations of the systems. This can be intuitively understood because the local dynamics of the system are effectively more decoupled, and thus additionally having, for example, smaller coupling constants between the first (input) qubit and the rest can significantly impact performance.

Fig.~\ref{fig:IPC}b illustrates the performance in the ergodic regime, where we do reach saturation of the maximum capacity for low noise (perfect saturation can never be observed due to numerical error, as explained in \onlinecite{Dambre}). 
In this regime, performance across different system realizations is much more homogeneous, and the effect of the different types of noise is more comparable, despite bit flip remaining the most disruptive. The latter effects also agree with the trends present in Fig.~\ref{fig:NMK_W0W10_noerrobars}.

We comment that in all cases the local observable set $Z$ presented a greater robustness to both kinds of noise (see Supplementary Figures~\ref{fig:IPC_W0_othersets}b and \ref{fig:IPC_SNR_comparison_Z} of the Appendix), which can be explained by the fact that the information stored in correlations dies out faster with decoherence than the one stored locally. Nonetheless, in the presented results we focus on the observable sets that provide a greater number of output functions than that of physical qubits, in order to unequivocally account for the exploitation of the greater dimensionality of the Hilbert space in the reservoir's performance.

Finally, we investigate in Fig.~\ref{fig:IPC}c the performance of the ergodic regime as a function of the nature of the observables. 
We first note that the normalized IPC does not differ qualitatively between the $ZZ$ and $Z+ZZ$ (larger) sets, reaching saturation in both cases (see the comparison in Appendix~\ref{appendix:IPC_other_observable_sets}). 
Furthermore, we compare the $XX + XY + YX + YY$ set, which corresponds to inherently quantum properties and is generally more cumbersome to measure in an experiment, to the time-multiplexed measurement of the $ZZ$ pairs. The time multiplexing scheme was introduced in \onlinecite{Fujii_Nakajima_2017} and consists of taking $V$ additional measurements during the transient regime, i.e., in $\Delta t/V$ intervals. We choose $V=4$ to equal the 40 elements of the $XX + XY + YX + YY$ set to the $10 \times V$ output functions of multiplexing the $ZZ$ set. We only consider bit flip noise for the multiplexed analysis, which is more detrimental than phase flip. The resulting comparison is presented in Fig.~\ref{fig:IPC}c, where we observe that the performance of both schemes is similar, but the multiplexed case presents itself as slightly more robust to noise. Although the difference is small in the small systems considered here, this result suggests that it may be more efficient to time-multiplex commuting observables, which are easier to measure, than to measure larger and more complicated sets of observables, at least in the system in Eq.~\eqref{eq:Ham_hW}.
%\mcs{{\large[[Figure 5 is one of the main results of this work, it deserves more detailed description.]]}}

\begin{figure}[htb]
    \centering
    \includegraphics[width = 0.45\textwidth]{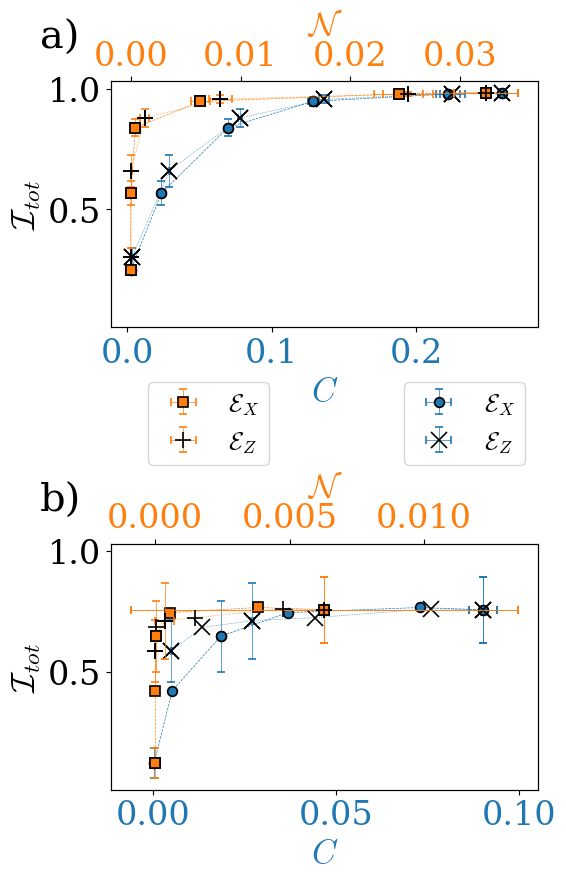}
    \caption{Total IPC for the $ZZ$ set vs $C$ and $\mathcal{N}$ for W=0 in (a) and W=10 in (b). The coherence and negativity values of the x-coordinates are shown in or can be extracted from Fig.~\ref{fig:NMK_W0W10_noerrobars}, and the y-coordinates from the total normalized capacity shown in Fig.~\ref{fig:IPC}a. (for $W=10$) and  Supplementary Figure~\ref{fig:IPC_W0_othersets}a, in Appendix~\ref{appendix:IPC_other_observable_sets} (for $W=0$), with each point in this plot corresponding to a different value of $p_{err}$. Limits on the $y$-axis are the same for both plots and error bars in $C$ are too small to be seen. In b), only some representative error bars are shown in order to aid visibility; the corresponding plot containing all error bars can be found in Appendix~\ref{appendix:additional_figs_performance_corrs} (Fig.~\ref{fig:fig6_all_error_bars}).}
    \label{fig:IPCvsX}
\end{figure}

\subsection{\label{sec:corr_vs_performance}Correlations vs. performance}

In this section we directly compare the total capacity $\mathcal{I}_{tot}$ with different features of the reservoir state. First, we consider the total coherence $C$ and the average negativity $\mathcal{N}$ in Fig.~\ref{fig:IPCvsX}. We highlight the smaller magnitude of the $x$-axes for $W=10$ with respect to $W=0$. This, together with the larger variability among system realizations in the MBL regime (discussed in the previous section) leads to the larger error bars in Fig.~\ref{fig:IPCvsX}b for $\mathcal{N}$.
Thus, even though the mean values for $W=10$ follow similar trends as those in $W=0$, we will limit the conclusions of this comparison to the ergodic regime, where this direct depiction confirms the monotonous relationship between correlations and performance. 
Our results are consistent with those in \onlinecite{xia_configured_2023}, where  this relationship is investigated by building a synthetic reservoir whose coherence can be exactly tuned by changing the input injection scheme. Within this setting, the authors find an increase in the performance of both STM (short-term memory) and PC (parity check) tasks with increasing coherence.
Our results also show that pure quantum correlations (i.e., entanglement) are not the main resource on which the performance of this reservoir is based, since $\mathcal{N}$ can drop by more than two-thirds from its noiseless value before the overall capacity is significantly decreased. On the other hand, the decay of the performance shows a stronger correlation with the loss of coherence. Different sets of observables similarly show that negativity has a weaker dependence on performance with respect to all other correlation indicators analyzed (see Appendix~\ref{appendix:additional_figs_performance_corrs}). These results suggest that this reservoir's performance relies more on the quantum phenomenon of superposition than on entanglement.
Interestingly, both bit flip and phase flip data seem to collapse under the same dependence, pointing towards the possibility that the relationship shown is a general property of local, Markovian noise such as the one described by both noise models. 

\begin{figure}[htb]
    \centering
    \includegraphics[width = 0.45\textwidth]{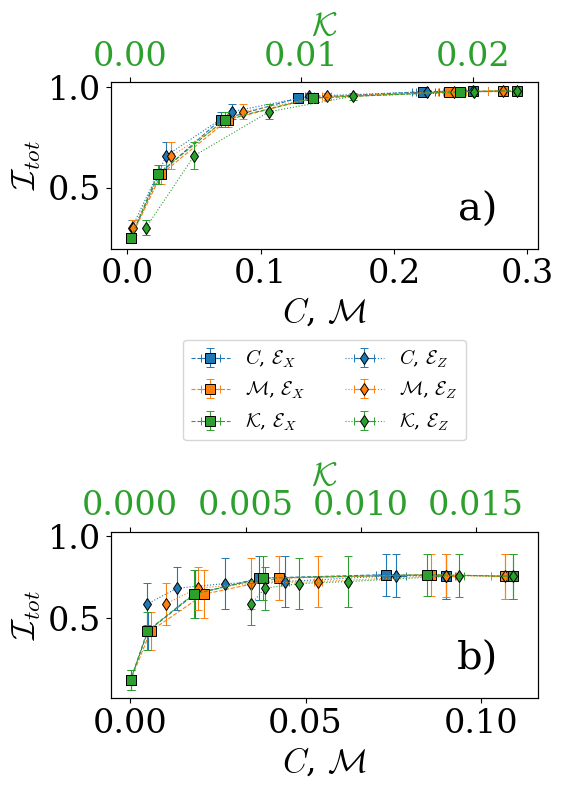}
    \caption{Total IPC for the $ZZ$ set vs $C, \ \mathcal{M}$ and $\mathcal{K}$ for W=0 in (a) and W=10 in (b). The coherence and negativity values of the x-coordinates are shown in or can be extracted from Fig.~\ref{fig:NMK_W0W10_noerrobars}, and the y-coordinates from the total normalized capacity shown in Fig.~\ref{fig:IPC}a. (for $W=10$) and  Supplementary Figure~\ref{fig:IPC_W0_othersets}a, in Appendix~\ref{appendix:IPC_other_observable_sets} (for $W=0$), with each point in this plot corresponding to a different value of $p_{err}$. Limits on the $y$-axis are the same for both plots and error bars in the $x$-axis are generally too small to be seen.}
    \label{fig:IPC_vs_CMK}
\end{figure}
The comparison of the IPC with the full coherence, the quantum hookup and the totally classical correlations is presented in Fig.~\ref{fig:IPC_vs_CMK}, where we present $C$ again in order to demonstrate the similarity between $C$ and $\mathcal{M}$ that was already mentioned in Section II.B.
From this data it is hard to specify the role of the totally classical correlations in an arbitrary case, but two observations can be pointed out: first, $\mathcal{K}$ follows a very similar trend as $\mathcal{M}$ (and, consequently, $C$) for both $W=0$ and $W=10$; and second, that their importance varies depending on the reservoir's dynamical regime. For instance, for very strong phase flip noise most of the remaining processing capacity for $W=10$ is due to the classical correlations $\mathcal{K}$. This is consistent with the results in Fig.~\ref{fig:NMK_W0W10_noerrobars}b and Fig.~\ref{fig:IPC_SNR_comparison_ZZ}b of Appendix~\ref{appendix:gaussian_smoothening}, suggesting that the reservoir heavily relies on completely classical resources in the localized phase.

\section{\label{sec:discussion}Discussion}

Our research has established a clear and consistent relationship between coherence and reservoir performance, as well as a notable link between the IPC of the reservoir and quantum correlations. 
Furthermore, we have identified quantum superposition and coherence, rather than entanglement, as the primary source of the (quantum) correlations that influence the reservoir performance in the described set-up (for a discussion on the effect of different input encodings, see Appendix~\ref{appendix:impact_input_encoding}).
%which our analysis reveals mostly have their origin in the phenomenon of quantum superposition rather than entanglement

An important point to emphasize is the critical impact of neglecting the signal-to-noise ratio (SNR) resulting from a finite number of measurements and uncontrolled error sources. This oversight can significantly impact the conclusions drawn from reservoir capacity studies. Therefore, beyond the necessary and often constructive role played by dissipation to provide fading memory  \cite{sannia2022dissipation,rodrigo2},  the inclusion of decoherence and noise needs to be addressed
accounting for statistical noise \cite{Govia_2021,mujal2023time,polloreno2023limits,hu2023tackling,nokkala2021high,PhysRevApplied.20.014051}. In particular, our results show that when realistic SNR is accounted for, there is no discernible improvement in reservoir performance in the presence of noise. This has been established in different regimes, suggesting that some non-monotonic dependencies reported in previous studies \cite{gotting2023exploring} won't hold in a more realistic setting.

In addition, our investigation of correlations has shed light on the fact that the reservoir, specifically set at the ergodic point under study ($W=0$), exploits quantum effects to enhance its performance. Our analysis also shows that in the considered MBL point $W=10$, 
purely classical correlations play a dominant role in maintaining the processing capacity of the reservoir.
These observations highlight the interplay between quantum dynamics and reservoir capabilities. In addition, our results suggest that the system's ability to build coherence can serve as a valuable indicator of the specific dynamical phase it is in. This insight contributes to a deeper understanding of the intricate quantum behaviors that influence reservoir dynamics.

\section{\label{sec:methods}Methods}

\subsection{\label{sec:corr_measures}Correlation measures}\label{sec:correlation_measures}
In this section we define the correlation indicators used throughout this work, as well as the details of their numerical simulation.
Following the order in which they are introduced, the coherence $C$ of a state $\rho$ is defined as:
\begin{equation}
    C(\rho) = \sum_{i\neq j} |\rho_{ij}|,
\end{equation}
referred to as $C_{l_1}$ in \onlinecite{colloquim_coherence}. Its maximal value is $2^N-1$, which is the factor by which we normalize it in Fig.~\ref{fig:model_heatmap_p0}b.

The total mutual information $\mathcal{T}(\rho)$, which expresses the total correlations of the system, is equivalent to the relative entropy between the state $\rho$ and the tensor product of its partial traces $\pi[\rho] = \pi_0[\rho] \otimes \pi_1[\rho] \otimes ... \otimes \pi_N[\rho] $, where $\pi_i[\rho] = \text{Tr}_i(\rho)$ \cite{modi_unified_2010}. As shown in Theorem 1 and proof of Lemma 1 of the same paper, the relative entropy between these two particular states can be described as:
\begin{equation}
    \mathcal{T}(\rho) = S(\rho||\pi[\rho]) = S(\pi[\rho]) - S(\rho),
    \label{eq:mutual_info}
\end{equation}
where $S(\cdot||\cdot)$ is the relative entropy and $S(\rho)= - \text{Tr}\left[\rho \log{\rho}\right]$ is the von Neumann entropy.
The quantum hookup $\mathcal{M}$, introduced in \onlinecite{giorgi_hallmarking_2018}, is the distance to the closest incoherent product (i.e., computationally useless) state, given by
\begin{equation}
    \mathcal{M}(\rho) = S(\Delta[\pi[\rho]] || \rho),
    \label{eq:quantum_hookup}
\end{equation}
where $\Delta[\rho] = \text{diag}(\rho)$ is the diagonal state obtained by applying the fully dephasing operation. As stated in \onlinecite{giorgi_hallmarking_2018}, both $\mathcal{T}$ and $\mathcal{M}$ are related by the local coherences $C_L = C(\pi[\rho])$ such that $\mathcal{M} = \mathcal{T} + C_L$; and since in our system the local coherences are always 0 (exactly in the ergodic case, and up to a $\sim 10^{-6}$ error in the MBL regime), the quantum hookup and the mutual information coincide.
In Fig.~\ref{fig:NMK_W0W10_noerrobars} we also plot the totally classical correlations $\mathcal{K}$, defined as 
\begin{equation}
    \mathcal{K}(\rho) = \mathcal{T}(\Delta[\rho]) ,
    \label{eq:totally_classical_correlations}
\end{equation}
and which quantify exclusively classical correlations, albeit not in their entirety.
As a measure for exclusively quantum correlations, we have analyzed the negativity $\mathcal{N}$ \cite{vidal2002} averaged over all possible partitions of the system, computed as follows:
\begin{equation}
    \mathcal{N}(\rho) = \mathbb{E}_A \left[ \frac{||\rho^{T_A}||_1 - 1}{2} \right],
    \label{eq:negativity}
\end{equation}
where $\rho^{T_A}$ is the partial transpose of $\rho$ with respect to subsystem $A$ and $||\cdot ||_1$ is the trace norm.

\subsection{\label{sec:IPC_calculation_details}Computation of IPC}

The explicit form of the contribution to the capacitance that stems from each separate task, see Eq.~\eqref{eq:targetfuns_IPC}, is given as 
\begin{equation}
    C_L = 1 - \frac{\min_{\{W\}} {MSE_L(y, \overline{y})}  }{\prom{\overline{y}^2}_L} ,
    \label{eq:def_capacity}
\end{equation}
where the $L$ indicates a long-time average over the input string of size $L$. $\prom{\overline{y}^2}_L$ is the square average of the target and $\min_{\{W\}} MSE_L(\cdot)$ is the mean squared error between the prediction produced from the trained weights $W$ and the target function, defined as follows: 
%%% commented representation with gather command because otherwise I lose numbered lines in between
% \begin{gather}
%     \prom{\overline{y}^2}_L = \frac{1}{L}\sum_{k=1}^L \overline{y}_k^2, \\
%     MSE_L(y, \overline{y}) = \frac{1}{L}\sum_{k=1}^L (y_k - \overline{y}_k)^2.
%     \label{eq:def_MSE}
% \end{gather}
\begin{equation}
    \prom{\overline{y}^2}_L = \frac{1}{L}\sum_{k=1}^L \overline{y}_k^2,
\end{equation}
\begin{equation}
    MSE_L(y, \overline{y}) = \frac{1}{L}\sum_{k=1}^L (y_k - \overline{y}_k)^2.
    \label{eq:def_MSE}
\end{equation}

We follow the order in which one should collect all these contributions so that they can be probed from largest to smallest from the Supplementary Material of the original formulation of the IPC \cite{Dambre}.
In practice, for the system size under consideration, we can reduce the upper bound to $d_{max}=6$, since the contributions become less significant as the degree is increased.

We also comment on the parameters of the data collection that were used for a stable calculation of the IPC:
after an initial number of time steps $\zeta=1000$ in which we let the system reach a stationary regime (the wash-out time), we collected training data for $L_t$ time steps and used the least squares method to obtain $N+1$ weights. These were then used to make predictions from the testing input string, of length $L_{uk}$. We set $L_t = L_{uk} = L$, although this is not a necessary ratio. Throughout this work, IPC calculations are done for $L = 10^5$ and $\Delta t=10$, which was observed to be sufficient to converge to a stationary regime. In addition, IPC averages are always taken over 10 different Hamiltonian realizations.

\section*{Data availability}

Data is available from the corresponding author at reasonable request.

%%%%%%%%%%%%%%%
\section*{Acknowledgements}
A.P. acknowledges the support of the European Commission FET-Open project AVaQus (GA 899561), the Agencia de Gestió d'Ajuts Universitaris i de Recerca through the DI grant (No. DI74) and the Spanish Ministry of Science and Innovation through the DI grant (No. DIN2020-011168).
%%
%%%%% From previous manuscript
This work was also supported by the Ministry for Digital Transformation and of Civil Service of the Spanish Government through the QUANTUM ENIA project call - Quantum Spain project, and by the European Union through the Recovery, Transformation and Resilience Plan - NextGenerationEU within the framework of the Digital Spain 2026 Agenda, the Spanish State Research Agency, through the María de Maeztu project CEX2021-001164-M funded by the MCIN/AEI/10.13039/501100011033,  and through the COQUSY project PID2022-140506NB-C21 and -C22 funded by MCIN/AEI/10.13039/501100011033.
G.L.G. is funded by the Spanish MEFP/MiU and co-funded by the University of the Balearic Islands through the Beatriz Galindo program (BG20/00085). R.M.P. acknowledges the QCDI project funded by the Spanish Government. The CSIC Interdisciplinary Thematic Platform (PTI) on Quantum Technologies in Spain is also acknowledged.
%%%%%%%%%%%%%%%
%%%%%%%%%%%%%%%

\section*{Author contributions}
%RZ and MCS defined the research topic and supervized the research. JN performed the numerical simulations. RMP, JN and GLG contributed to the theoretical aspects of the paper (demonstrations of Lemma 1 and Universality theorem). VP assessed the potential experimental implementation.
AP coded and performed the numerical simulations.
% except for Fig.~\ref{fig:model_heatmap_p0}a., which was generated by R.M..
All authors contributed to interpreting the results and writing the article.

\section*{Competing interests}

We declare no competing interests.

\bibliography{bibliography.bib}
\bibliographystyle{naturemag}

%%%%%%%%%% Merge with supplemental materials %%%%%%%%%%
\pagebreak
\widetext
% \onecolumngrid
% \newpage
% \vspace{0.5cm}
% \begin{center}
%     {\large \textbf{ Supplementary Information for Role of coherence in many-body Quantum Reservoir Computing}}
% \end{center}
%%%%%%%%%% Merge with supplemental materials %%%%%%%%%%
%%%%%%%%%% Prefix a "S" to all equations, figures, tables and reset the counter %%%%%%%%%%
\setcounter{equation}{0}
\setcounter{figure}{0}
\setcounter{table}{0}
\setcounter{page}{1}
\setcounter{section}{0}
% \makeatletter
\renewcommand{\theequation}{S\arabic{equation}}
\renewcommand{\thefigure}{S\arabic{figure}}

\appendix

\section{\label{appendix:trotterization_lindblad}Trotterization of Lindblad evolution and implementation}

The single-qubit bit flip and phase flip channels are given in \cite{NielsenChuang} as
\begin{equation}
    \text{Phase flip: } \mathcal{E}_{z}[\rho] = (1-p) \rho + p\ \sigma_z \rho \sigma_z 
    \label{phaseflip_ch}
\end{equation}
\begin{equation}
    \text{Bit flip: } \mathcal{E}_{x}[\rho] = (1-p) \rho + p\ \sigma_x \rho \sigma_x
    \label{bitflip_ch}
\end{equation}
where $p=p_{err}$ is the error probability.
We extend the previous scenario to multi-qubit systems by considering that dissipation happens independently in all qubits, following previous literature on the subject \cite{ChenNurdin}. In this protocol, we implement $\eta$ decoherence steps between inputs, where on each decoherence step we first let the system evolve unitarily for a time $\delta t = \Delta t / \eta$ and then apply the following dissipative map:
\begin{equation}
    \mathcal{E}_{\alpha}[\rho] = (1-p)^N \rho + (1-p)^{N-1} p\sum_i
   \sigma_\alpha^i \rho \sigma_\alpha^i +
   (1-p)^{N-2}p^2  \sum_{i \neq j}
   \sigma_\alpha^i \sigma_\alpha^j \rho \sigma_\alpha^i \sigma_\alpha^j + ... +
   p^N \sigma_\alpha^1 ... \sigma_\alpha^N \rho \sigma_\alpha^1 ... \sigma_\alpha^N
    \label{eq:dissipative_map}
\end{equation}
where $\alpha = \{z,x\}$  and $i,j = 1,..., N$, so that $\sigma_\alpha^i$ corresponds to the Pauli matrix of coordinate $\alpha$ acting on spin $i$. In our simulations we fixed $\eta=50$.

The previous alternation between unitary maps and decoherence steps is equivalent to a generalized Trotterization \cite{han2021experimental}:
\begin{equation}
    \lim_{\eta\rightarrow \infty}\left(e^{\mathcal{H}\frac{\Delta t}{\eta}} \prod_j e^{\mathcal{L}_j\frac{\Delta t}{\eta}}\right)^{\eta}=e^{(\mathcal{H}+\sum_i\mathcal{L}_i)\Delta t},
\end{equation} 
where $e^{\mathcal{H}\frac{\Delta t}{\eta}}$ is the corresponding superoperator of the unitary dynamics. Notice that the superoperators $e^{\mathcal{L}_j\frac{\Delta t}{\eta}}$ with Lindblad equation
\begin{equation}
    \dot{\rho}=\gamma\sigma_\alpha\rho\sigma_\alpha-\gamma\rho
\end{equation} 
are equivalent to the quantum channels of Eqs.~\eqref{phaseflip_ch} and \eqref{bitflip_ch}. Therefore, the quantum reservoir dynamics can be approximated by
\begin{equation}
    \rho[(k + 1)\Delta t] \simeq e^{(\mathcal{H}+\sum_i\mathcal{L}_i)\Delta t} \left(\rho_1 \otimes \text{Tr}_1\rho[k\Delta t]\right),
    \label{eq:QRC_lindblad}
\end{equation} 
where the superoperator $e^{(\mathcal{H}+\sum_i\mathcal{L}_i)\Delta t}$ is given by the Lindblad equation in Eq. (3) of the main text.

\section{\label{appendix:IPC_other_observable_sets}IPC for different sets of observables}

\begin{figure}[h!]
    \centering
        \includegraphics[width=0.9\textwidth]{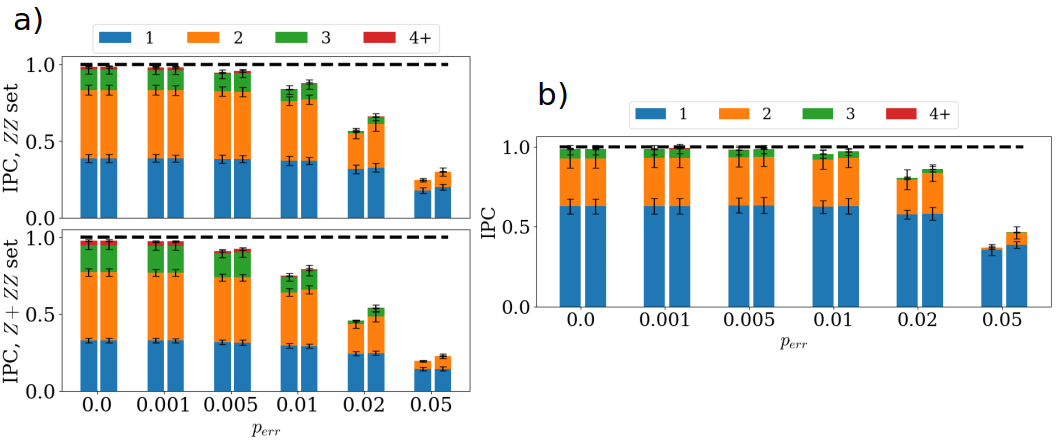}
    \caption{IPC contributions for $W=0$ for the a) $ZZ$ (top) and $Z+ZZ$ observable sets (bottom) and b) the $Z$ observable set. Left and right bars correspond to $\mathcal{E}_X$ and $\mathcal{E}_Z$ noise, respectively.}
    \label{fig:IPC_W0_othersets}
\end{figure}

We present in Fig. \ref{fig:IPC_W0_othersets} the results calculated from sets of observables different from the ones in the main text, where signal-to-noise ratio (SNR) has been considered, namely a statistical noise of $\overline{\sigma}=0.001$. Averages are taken over 10 Hamiltonian realizations.

\section{\label{appendix:gaussian_smoothening}Effect of statistical noise}

Following \cite{Fujii_Nakajima_2017}, we collect the output functions considering ensemble measurements, such that we disregard the effect of back-action.
Thus, observables are computed as 
\begin{equation}
    x_i(k\Delta t) = \left\langle\sz_i\right\rangle = \text{Tr}\left[\sz_i \rho (k\Delta t)\right]
    \label{eq:measurement}
\end{equation}
However, in a real experimental set-up, precision is limited by the finite number of averaging shots and the underlying noise sources of the measurement procedure \cite{mujal2023time,PhysRevApplied.20.014051,khan2021physical}.
To approach realistic experimental results, the 12-digit precision of the observable quantities from Eq.~\eqref{eq:measurement} given by numerical simulation is mitigated by adding noise from a normal distribution of mean $\mu=0$ and standard deviation $\overline{\sigma} = 0.001$, such that:
\begin{equation}
    \Tilde{x}_i (k \Delta t) = x_i(k\Delta t) + n_i, \quad n_i \sim N(\mu, \overline{\sigma}^2)
    \label{eq:smoothened_measurement}
\end{equation}

\begin{figure}[h]
    \centering
    \includegraphics[width=\textwidth]{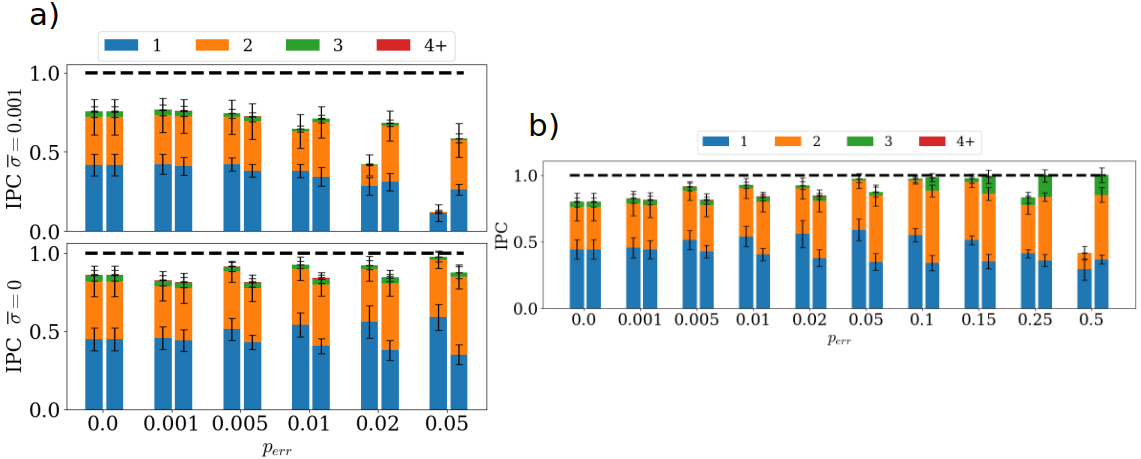}
    \caption{IPC for $W=10$ and the $ZZ$ set. a) shows the comparison of the IPC with and without considering statistical noise, while b) presents the IPC for an infinitely large number of ideal measurements for a larger range of $p_{err}$. The left and right bars correspond to $\mathcal{E}_X$ and $\mathcal{E}_Z$ noise, respectively.}
    \label{fig:IPC_SNR_comparison_ZZ}
\end{figure}

\begin{figure}[h]
    \centering
    \includegraphics[width=0.8\textwidth]{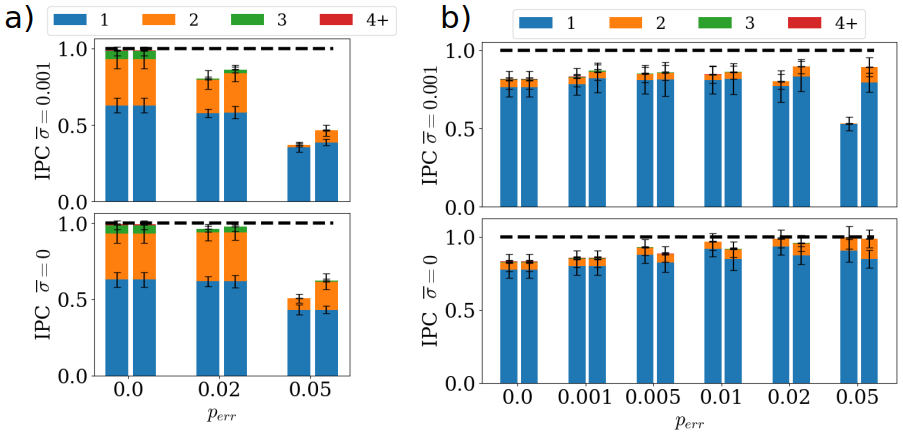}
    \caption{IPC contributions for the $Z$ set with (top) and without (bottom) considering statistical noise for a) $W=0$ and b) $W=10$. The left and right bars correspond to $\mathcal{E}_X$ and $\mathcal{E}_Z$ noise, respectively. In a), IPC for $p_{err}=0.001, 0.005, 0.01$ is omitted because it's very similar to the ideal case in both scenarios.}
    \label{fig:IPC_SNR_comparison_Z}
\end{figure}

As highlighted in the main text, we confirmed that the absence of this smoothening procedure can lead to misleading interpretations of the results, since a real experimental setup will always be subject to statistical noise. In Figs.~\ref{fig:IPC_SNR_comparison_ZZ} and \ref{fig:IPC_SNR_comparison_Z} we exemplify this by plotting the resulting IPC in the MBL regime (and in the ergodic one in Fig.~\ref{fig:IPC_SNR_comparison_Z}a) with and without considering a finite number of measurements, always averaging over 10 Hamiltonian realizations. Notice that in Figs.~\ref{fig:IPC_SNR_comparison_ZZ}a and \ref{fig:IPC_SNR_comparison_Z}b, while in the case where shot noise is taken into account noise strength anticorrelates with performance, in the case where we preserve simulation precision the interpretation would be the opposite one. In addition, Fig.~\ref{fig:IPC_SNR_comparison_Z}a shows that considering simulation precision alone leads to unrealistic robustness to phase flip and bit flip noises. 
Finally, Fig.~\ref{fig:IPC_SNR_comparison_ZZ}b shows that even with simulation precision, bit flip noise eventually kicks in in the processing capacity of the system, completely destroying it for maximum noise strength ($p_{err}=0.5$). In contrast, the capacity remains maximal for maximum phase flip noise strength. This further reinforces the observation that the MBL phase stores its correlations along the $Z$-direction, as presented and discussed in the main text. Furthermore, considering these results together with those in Fig. 3b of the main text, one can conclude that the capacity found for strong phase flip noise in the MBL phase is due to totally classical correlations in a relatively large part (which at the strongest noise point considered, $p_{err}=0.05$, constitute around 50\% of the total correlations).

\section{\label{appendix:additional_figs_performance_corrs}Additional figures on the comparison of performance vs. correlations}

\begin{figure}[h!]
    \centering
    \includegraphics[width=\textwidth]{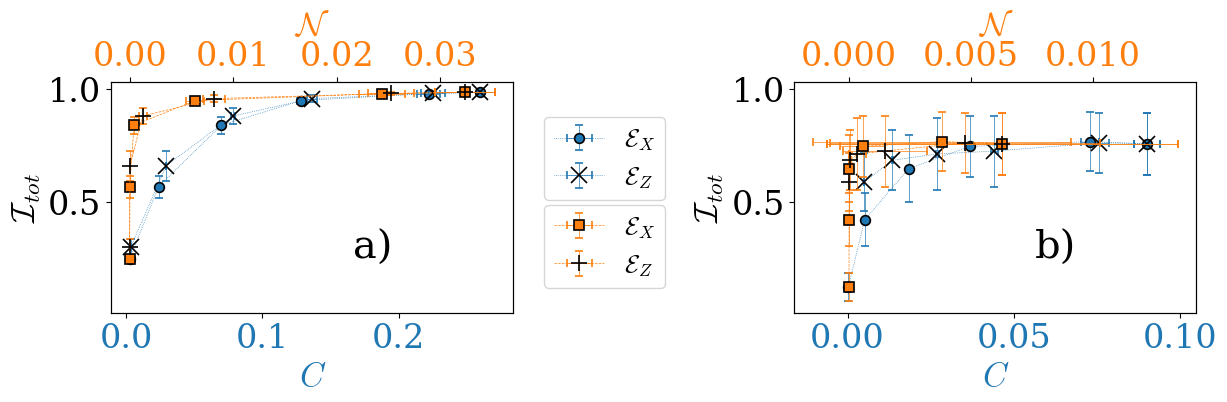}
    \caption{Total IPC for the $ZZ$ set vs $C$ and $\mathcal{N}$ for W=0 in (a) and W=10 in (b). The coherence and negativity values of the x-coordinates are shown or can be extracted from Fig. 3 of the main text, and the y-coordinates form the total normalized capacity shown in Fig. 5a of the main text (for $W=10$) and Figure~\ref{fig:IPC_W0_othersets}a (for $W=0$), with each point in this plot corresponding to a different value of $p_{err}$. Limits on the $y$-axis are the same for both plots and error bars in $C$ are too small to be seen.}
    \label{fig:fig6_all_error_bars}
\end{figure}

\begin{figure}[h!]
    \centering
    \includegraphics[width=0.8\textwidth]{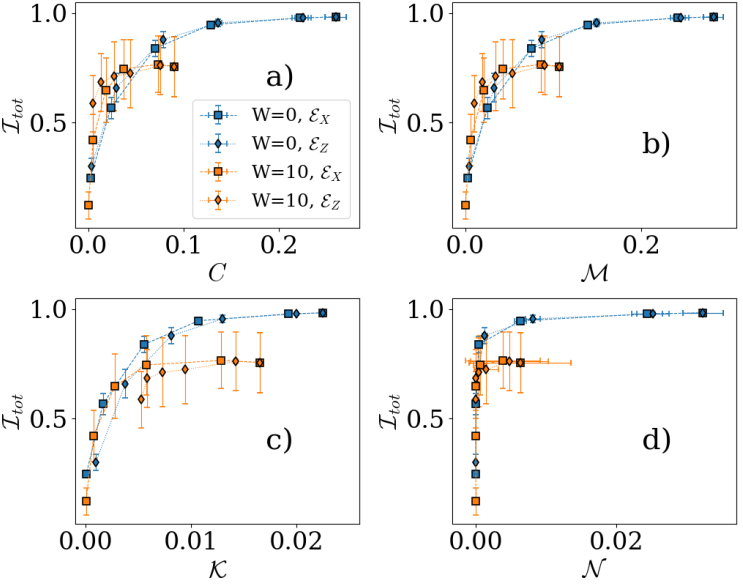}
    \caption{Total IPC for the ZZ observable set vs. coherence (a), quantum hookup (which in this system is equal to the total correlations up to $\sim 10^{-6}$) (b), totally classical correlations (c) and negativity (d) for all the regimes and noise types under study.}
    \label{fig:IPCvs_C_M_K_N}
\end{figure}

\begin{figure}[h!]
    \centering
    \includegraphics[width=0.8\textwidth]{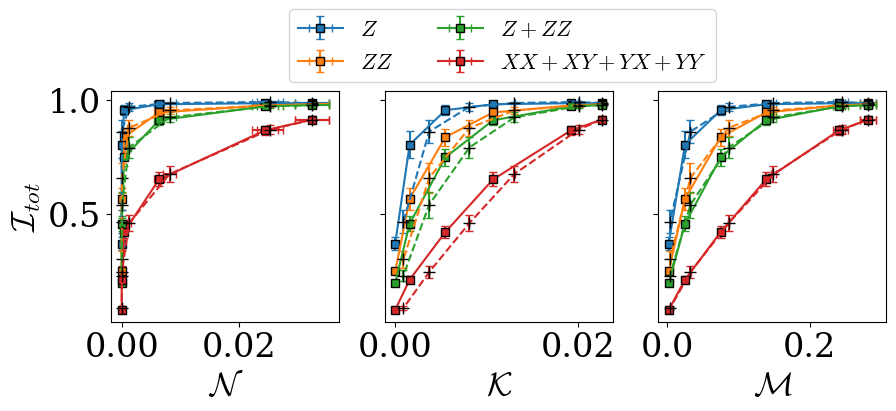}
    \caption{Total IPC vs $\mathcal{N}, \mathcal{K}$ and $\mathcal{M}$ for different sets of observables with increasing contributions from nonlinear capacities in the ergodic regime ($W=0$) and considering $\overline{\sigma}=0.001$ for various sets of observables. Dashed lines and crosses correspond to phase flip noise while squares and solid lines refer to bit flip noise data.}
    \label{fig:Itot_vs_NKM}
\end{figure}

In this section we provide some complementary figures to Section II.E of the main text.
Firstly, we include the plot presented in Fig. 6 of the main text including all the error bars in a larger format in Fig.~\ref{fig:fig6_all_error_bars} of this supplementary document, for completeness.
Secondly, we provide individual comparisons of the performance with each of the correlation measures under study in the main text. As it can be seen comparing Figs.~\ref{fig:IPCvs_C_M_K_N}a and \ref{fig:IPCvs_C_M_K_N}b, coherence and quantum hookup provide very similar information. This is due to the relation $\mathcal{M} = C + \mathcal{K}$ \cite{giorgi_hallmarking_2018} and the smaller scale of the totally classical correlations (notice the difference of the scale of the x-axis between the aforementioned plots and Fig.~\ref{fig:IPCvs_C_M_K_N}c). Apart from the more distinguishable tendencies of the two kinds of noise in the MBL phase, the totally classical correlations also follow the general trend of $C$ and $\mathcal{M}$, leaving only the negativity (in Fig.~\ref{fig:IPCvs_C_M_K_N}d) with a markedly different behavior.
Lastly, we extend here the comparison of performance vs. correlations to other observables, which further reinforces the generality of the conclusions derived for the ergodic regime in the main text. Figure~\ref{fig:Itot_vs_NKM} indeed shows the stronger dependence of performance on $\mathcal{K}$ and $\mathcal{M}$ (and consequently, on $C=\mathcal{M}-\mathcal{K}$) in comparison to the dependence with $\mathcal{N}$ for all observables. We attribute the stronger decay within different types of observable sets to the fact that most sensitive sets have higher nonlinear processing capacities, which are less robust to the noise model under consideration. This consequently implies a stronger dependence on all the correlation measures under study.


\section{\label{appendix:impact_input_encoding}Impact of input encoding}

\begin{figure}
    \centering
    \includegraphics[width=0.85\textwidth]{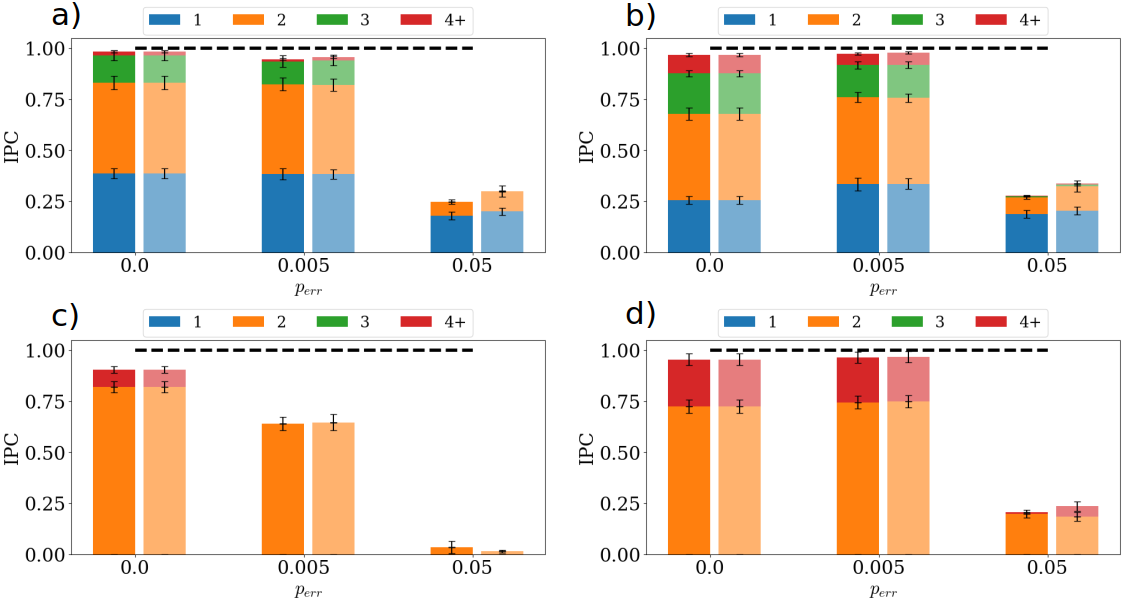}
    \caption{IPC contributions in the ergodic phase for the $ZZ$ observable set, considering $\sim 10^6$ measurements. a) shows the encoding \eqref{eq:mixed_z} (mixed in z), b) corresponds to the encoding \eqref{eq:pure_zx} (pure, centered around z), c) to the encoding in \eqref{eq:mixed_x} (mixed in x) and d) to the encoding in \eqref{eq:pure_x} (pure, centered around x). In all cases, the left bar corresponds to bit flip and the right one to phase flip noise.}
    \label{fig:IPC_W0_diffent_encodings}
\end{figure}

In this section we investigate the effect of the choice of the input injection scheme on the results of the main text. As already studied in \cite{mujal2021analytical}, this choice can also impact the reservoir's information processing capabilities.
For the study of the additional encodings, we focus on the ergodic regime and three noise intensities ($p=0, 0.005$ and 0.05) for the sake of clarity. The encoding of the main text, which we will refer to as mixed encoding in the z-direction, is reproduced here in order to ease the comparison:
\begin{equation}
    \rho_1^{mixed, z} (s) = \begin{pmatrix}
        1-s & 0 \\
        0 & s
    \end{pmatrix} = (1-s) \ket{0} \bra{0} + s \ket{1}\bra{1}
    \label{eq:mixed_z}
\end{equation}
with $s \in [0, 1]$. Some of the previously shown results for this encoding are collected in Fig.~\ref{fig:IPC_W0_diffent_encodings}a in order to ease the comparison with the additional input encodings.

The first additional encoding we examine consists of injecting the input in a pure state along the x- and z-directions of the Bloch sphere
\begin{equation}
    \rho_1^{pure, z} (s) = \frac{1}{2}\begin{pmatrix}
        1 + \cos{(\pi s)} & \sin{(\pi s)} \\
        \sin{(\pi s)} & 1 - \cos{(\pi s)}
    \end{pmatrix} = \ket{\eta_+(s)} \bra{\eta_+(s)}
    \label{eq:pure_zx}
\end{equation}
where $\ket{\eta_+(s)} = \cos{\left(\frac{\pi s}{2}\right)} \ket{0} + \sin{\left(\frac{\pi s}{2}\right)} \ket{1}$. This input injection scheme introduces some coherence for inputs $s \in (0, 1)$, but the encoding scheme remains centered along the z-axis. As it can be shown by comparing Figs.~\ref{fig:IPC_W0_diffent_encodings}a and \ref{fig:IPC_W0_diffent_encodings}b, this encoding provides very similar robustness to the one in the main text (if not slightly higher), with some higher nonlinear processing capacity. 

\begin{figure}
    \centering
    \includegraphics[width=0.55\textwidth]{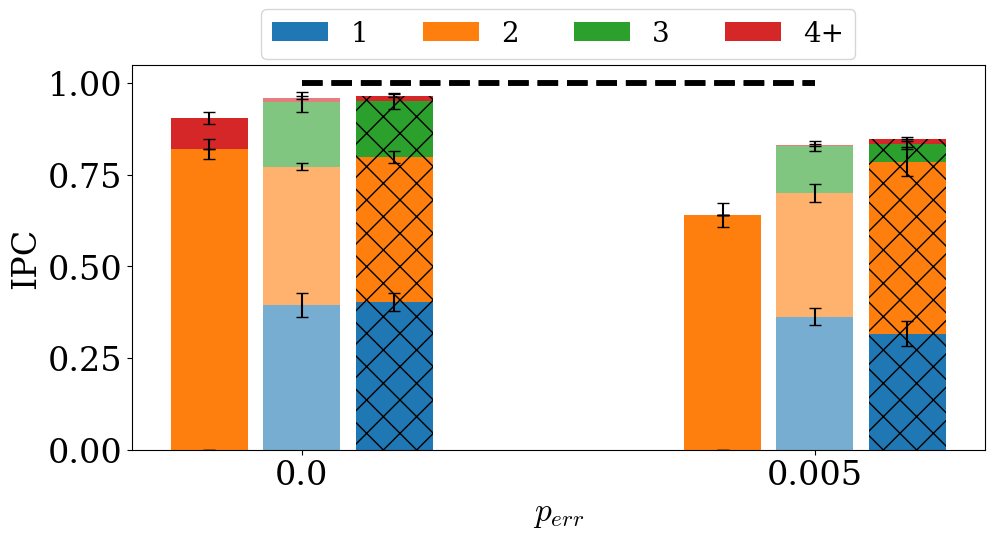}
    \caption{IPC contributions for the mixed encoding in x (see Eq.~\eqref{eq:mixed_x}) for $W=0$ in the noiseless and mild bit flip noise cases, statistical noise of $\overline{\sigma} = 10^{-3}$ and for the $ZZ$ observable set (left and right bars) and the $ZZ+ZX$ set (central bars). The right bars consider a system Hamiltonian with broken parity symmetry $H^\prime = \sum_{i, j}J_{ij} \sx_i \sx_j + \sum_i (\sz_i + 0.05 \sx_i)$, while the left and central ones consider the usual symmetric Hamiltonian \eqref{eq:ham_symmetric}.}
    \label{fig:IPC_mixedx_all}
\end{figure}

Secondly, we study the encoding analogous to \eqref{eq:mixed_z} along the x-direction, namely:
\begin{equation}
    \rho_1^{mixed, x} (s) = \begin{pmatrix}
        \frac{1}{2} & s - \frac{1}{2} \\
        s - \frac{1}{2} & \frac{1}{2}
    \end{pmatrix} = (1-s) \ket{-} \bra{-} + s \ket{+}\bra{+}
    \label{eq:mixed_x}
\end{equation}
Despite the similarity to \eqref{eq:mixed_z}, we observe remarkably different behaviors of the processing capacity with respect to the previous two encodings, as may be seen comparing Figs.~\ref{fig:IPC_W0_diffent_encodings}a and ~\ref{fig:IPC_W0_diffent_encodings}b with Fig.~\ref{fig:IPC_W0_diffent_encodings}c. Two features of the latter stand out: the participation in the IPC of even degrees only when taking the $ZZ$ set (same observables used for the previous encodings) and the reduced robustness to both kinds of noise. The observation of only even IPC degrees can be attributed to the parity symmetry of the Hamiltonian $H$ along the x-direction. We recall that 
\begin{equation}
    H = \sum_{i, j}J_{ij} \sx_i \sx_j + \sum_i \sz_i
    \label{eq:ham_symmetric}
\end{equation}
This symmetry imposes some limitations on the mixing among the $ZZ$ correlations of the information entering in the x-direction. However, if we consider a set of 10 observables including both $ZZ$ and $ZX$-type correlators (which we will refer to as the $ZZ+ZX$ set), the odd degree contributions to the IPC are restored and the robustness is slightly higher (see Fig.~\ref{fig:IPC_mixedx_all}, central bars). Since the choice of 10 observables of the $ZZ$ and $ZX$ type is not unique, we comment that we also looked into 10 randomly selected 2-qubit correlators among all 90 possible ones and obtained essentially identical results, so that we can be sure that the particular choice within the $ZZ+ZX$ set did not matter. In addition, similar results can be obtained directly with the $ZZ$ observable set if we break the symmetry of the Hamiltonian; this effect is shown in Fig.~\ref{fig:IPC_mixedx_all} (right bars, hatched), where the reservoir's natural dynamics was fixed to $H^\prime = \sum_{i, j}J_{ij} \sx_i \sx_j + \sum_i (\sz_i + 0.05 \sx_i)$.
The remaining difference in robustness even after the recovery of the odd contributions to the IPC thus seems to be due to the interplay between the input encoding and the Hamiltonian-induced dynamics.
In order to investigate this hypothesis further we study an additional encoding similar to \eqref{eq:pure_zx} but now centered around the x direction, i.e.,
\begin{equation}
    \rho_1^{pure, x} (s) = \frac{1}{2}\begin{pmatrix}
        1 + \sin{(\pi s)} & -\cos{(\pi s)} \\
        -\cos{(\pi s)} & 1 - \sin{(\pi s)}
    \end{pmatrix} = \ket{\xi_+(s)} \bra{\xi_+(s)}
    \label{eq:pure_x}
\end{equation}
where $\ket{\xi_+(s)} = \cos{\left(\frac{\pi s}{2}\right)} \ket{-} + \sin{\left(\frac{\pi s}{2}\right)} \ket{+}$. With this input encoding, now containing some projection of the input onto the z-axis, we observe in Fig.~\ref{fig:IPC_W0_diffent_encodings}d that robustness at mild noise intensity ($p_{err}=0.005$) is restored.

\begin{figure}
    \centering
    \includegraphics[width=\textwidth]{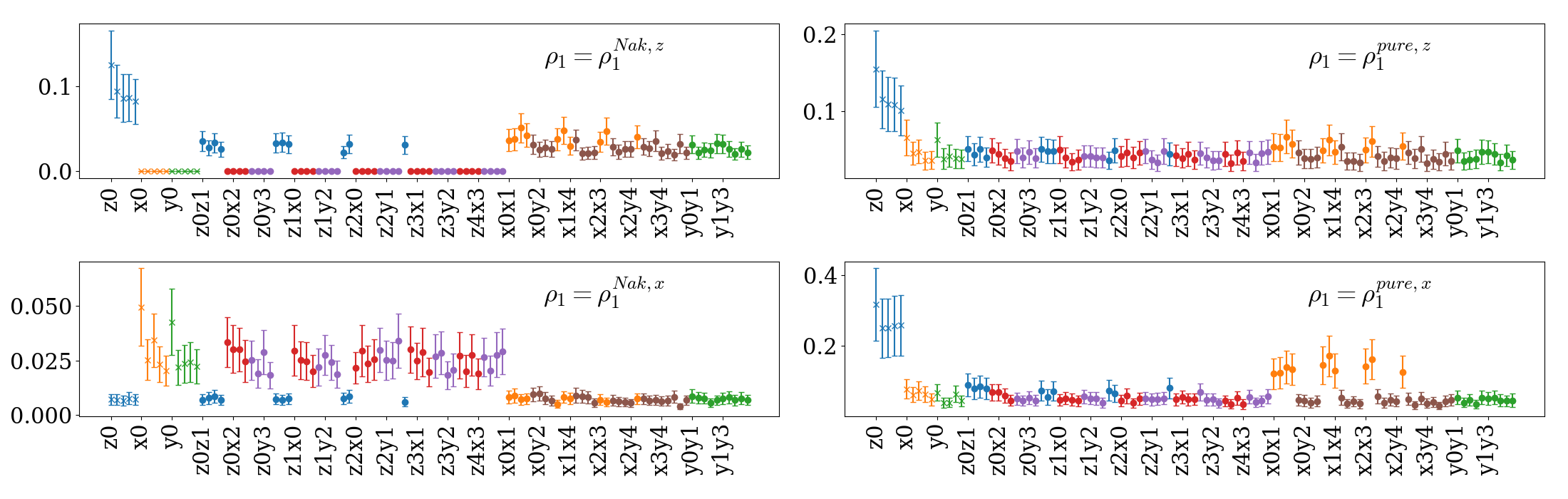}
    \caption{Average of the absolute value of local observables and all 2-local correlators in the stationary state, all subject to the same input stream in all cases. Crosses correspond to local observables and dots to correlations, and the color code is as follows: blue for observables only involving $Z$, orange for $X$, green for $Y$, red for type $ZX$, purple for type $ZY$ and brown for type $XY$. The data presented corresponds to the noiseless case and the ergodic regime, where averages are taken over 7000 data input steps after discarding the initial 1000 ones for wash-out and over the 10 systems under consideration.}
    \label{fig:loc_nonloc_obs_stationary_sate}
\end{figure}

Given the previous IPC results, it is interesting to look into the general features of the system in the stationary regime for each encoding in order to understand their performance.
To this end we present Fig.~\ref{fig:loc_nonloc_obs_stationary_sate}, which shows evidence that the information coming from the input is stored locally along the z-axis in the most robust encodings, leading to larger error bars in the stationary state for the $\prom{Z_i}$ observables. This is in contrast with the mixed encoding in x, where individual spins are mostly oriented along the x- and y-axes, but the signal coming from all these correlators and local quantities is much lower (notice the difference between y-scales). Overall, this points to this input injection scheme inducing a very small response in the reservoir, thus making it more sensitive to shot noise as well. Indeed, as shown in \cite{mujal2023time}, observables can be more sensitive to shot noise the higher the order of the Pauli string they involve. This makes the bottom left panel in Fig.~\ref{fig:loc_nonloc_obs_stationary_sate} consistent with the lower robustness of the encoding mixed in x, evidenced in Fig.~\ref{fig:IPC_W0_diffent_encodings}.

\begin{figure}
    \centering
    \includegraphics[width=0.7\textwidth]{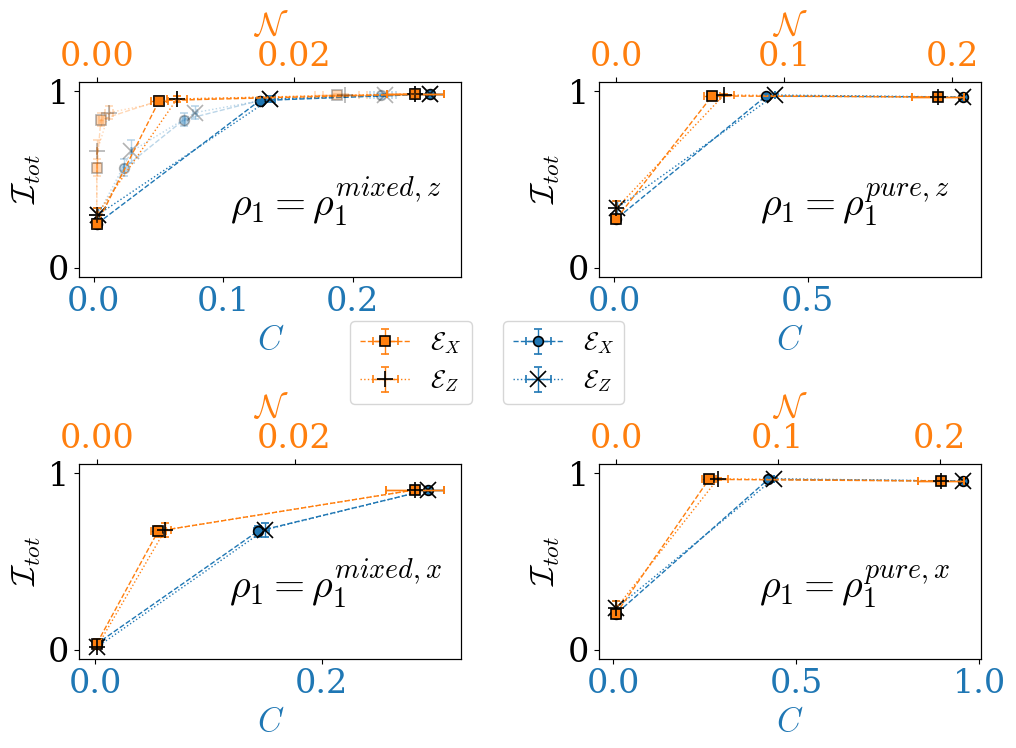}
    \caption{Correlations vs. total IPC for all encodings examined in the ergodic phase and with respect to the $ZZ$ observable set. The lighter curves in the top left plot correspond to all the data presented in Fig. 6 of the main text, whereas only the points corresponding to $p_{err}=0, 0.005, 0.05$ (the ones examined in the rest of the encodings) are left in the opaque curves.}
    \label{fig:IPCvsCvsN_all_encodings}
\end{figure}

To conclude the analysis, we present in Fig.~\ref{fig:IPCvsCvsN_all_encodings} the analogous comparison of IPC vs correlations of Fig. 6 of the main text. Despite the reduced set of points under consideration, this study suggests that the conclusion that the transverse-field Ising reservoir backs its performance in superposition rather than entanglement can be extended to a wide family of input encoding schemes. Our results point to this family encompassing those encodings that contain some projection of the input into the z-axis when considering input injection on a single qubit, since according to our analysis this allows for the system to be significantly modified by the input string, its performance staying robust while its entanglement drops.

\end{document}